# Scaling laws for the critical current density in anisotropic biaxial superconductors

Yingxu Li[1], Guozheng Kang[1] and Yuanwen Gao[2,3]

[1] Department of Engineering Mechanics, School of Mechanics and Engineering, Southwest Jiaotong University, Chengdu, Sichuan 610031, People's Republic of China

[2] Key Laboratory of Mechanics on Environment and Disaster in Western China, The Ministry of Education of China, Lanzhou, Gansu 730000, People's Republic of China

[3] Department of Mechanics and Engineering Science, College of Civil Engineering and Mechanics, Lanzhou University, Lanzhou, Gansu 730000, People's Republic of China

E-mail: ywgao@lzu.edu.cn and yingxuli@swjtu.edu.cn

**Abstract**: To understand the anisotropy of flux pinning and critical current density in technological superconductors, the scaling law for the anisotropy of single-vortex collective pinning in uniaxial superconductors is extended to flux-bundle collective pinning in biaxial superconductors. The scaling results show that in a system of random uncorrected point defects, the critical current density is described by a unified function with the magnetic field of the scaled isotropic superconductor. The obtained angular dependence of the critical current density depicts the main features of experimental observations, considering possible corrections due to the strong-pinning interaction.

## 1. Introduction

High-temperature superconductors (HTS) show an extensive application prospect in large-power magnets and cables, due to its relatively high critical temperature. How to improve the vortex-pinning properties for raising the critical current density and upper critical field, is a huge challenge in the area of high power application. An outstanding feature of HTS is the anisotropy of superconducting condensate, which manifests at the inequality of the microscopic superconductivity parameters or the upper critical field [1] along the crystallographic axes. For example, in single-crystal $NdFeAsO_{1-x}F_x$, the ratio of zero-temperature coherence lengths perpendicular to the FeAs layers and in the layers is ~4 [2]. In $(Ba,K)Fe_2As_2$ and $Nd(F,O)FeAs$ the anisotropy factors are ~2.5 and 7.5 [3], respectively. The field and angular dependences of the critical current density $J_c$ shed light on the anisotropy. Multiple techniques have been used to measure $J_c$ variation with the magnetic field direction in various superconducting samples [4-9]. The critical current density is fundamentally caused by the interaction of the flux vortices and defects [10]. To this end, the anisotropic $J_c$ depends not only on the anisotropy of superconducting condensate itself but on the defective nature. In high-temperature cuprate superconductors, columnar and planar defects exhibit a notably different field dependence of $J_c$ [8, 11-13]. Furthermore, the anisotropy in $J_c$ depends on the strength of the pinning interaction. From the observations on the behaviors of $J_c$ anisotropy, a strong-pinning interaction is demonstrated in the hole-doped $Ba_{0.6}K_{0.4}Fe_2As_2$ single crystal [9], whereas there is a weak-collective-pinning interaction after introducing point pinning defects.

A traditional way to incorporate the anisotropy into the phenomenological description of superconductivity is to introduce an anisotropic effective-mass tensor into the Ginzburg-Landau (GL) equations [14]. One then has to repeat all the calculations that have been done for the isotropic case before. A more elegant approach is the scaling law [14, 15], which scales the anisotropic problem to a corresponding isotropic one at the initial level of GL free energy. Reusing the scaling law, the isotropic results are then simply generalized to the anisotropic ones. Despite its effectiveness, the scaling rules are limited to treat the anisotropic uniaxial superconductors. On the other hand, single-vortex pinning receives the most concerns in the scaling rules, without considering the magnetic field dependence. Here, inspired by most recent findings on the biaxial anisotropy [16-19] (axes $a$, $b$ and $c$) and on the field dependence of the critical current density [8, 20-22], we extend the scaling rules to account for the anisotropic biaxial superconductors exposed to magnetic field with arbitrary direction and magnitude. Through this paper, a simple theory is established to unfold the complex physics in the anisotropy of the most concerned HTS. In the next section we first give the procedure for deducing the scaling rules for anisotropic biaxial superconductors within the single-vortex pinning regime, in light of those for uniaxial superconductor [14].

## 2. Scaling laws for anisotropic biaxial superconductors

### 2.1 Single-vortex pinning regime

The GL free-energy functional for the Gibbs free energy per unit volume is [14, 23]

$$f_s = f_n(0) + \alpha |\Psi|^2 + \frac{\beta}{2}|\Psi|^4 + \sum_{j=1}^{3}\frac{1}{2m_j}\left|\left(-i\hbar\frac{d}{dx_j} - 2e A_j\right)\Psi\right|^2 + \frac{B^2}{2\mu_0} - \mathbf{B}\cdot\mathbf{H}, \quad (1)$$

where $\Psi(\mathbf{r})$ is the order parameter, $\mathbf{A}$ is the magnetic vector potential, $\mathbf{B} = \nabla\times\mathbf{A}$ is the local magnetic induction strength, and $\mathbf{H}$ is the magnetic field. $f_n(0)$ is the free energy of the normal state at zero magnetic field, $B^2/2\mu_0$ is magnetic field energy and $-\mathbf{B}\cdot\mathbf{H}$ is the diamagnetic energy. The GL parameter $\alpha(T) = -\alpha(0)(1-T/T_c)$ changes sign at the critical temperature $T_c$, whereas $\beta$ is taken to be constant with respect to temperature. $e(>0)$ is the elementary charge. $m_j$, $j=1,2,3$, denote the effective masses along the principal axes of the crystal.

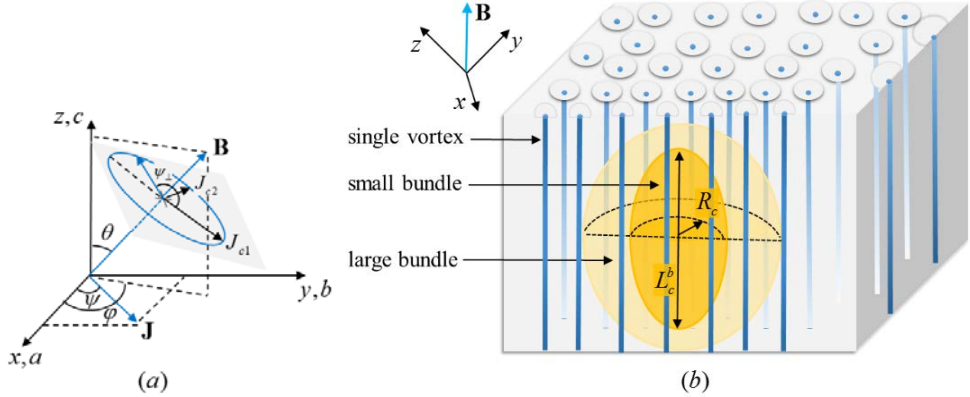

Figure 1. Schematic of the collective pinning model in an anisotropic biaxial superconductor. (a) A superconductor subject to the magnetic induction $\mathbf{B}$ and current $\mathbf{J}$. The direction of $\mathbf{B}$ is determined by the angles $\varphi$ and $\theta$, and angle $\psi$ specifies the direction of $\mathbf{J}$. The highlighted plane represents the plane perpendicular to $\mathbf{B}$, and $\mathbf{J}_{c\perp}$ is the projection of $\mathbf{J}$ in this plane, with $\psi_\perp$ representing the direction of $\mathbf{J}$ with respect to the principal axes of $\mathbf{J}_{c\perp}(\psi_\perp)$. (b) Collective pinning mechanism. Increasing the magnetic induction, the collective pinning successively acts on the single vortex, the small bundle and large bundle.

For simplicity and because HTS are within high accuracy biaxial (axes $\parallel c$, $\parallel a$ and $\parallel b$) materials, we denote the mass anisotropy ratio by $\varepsilon^2 = m_{ab}/m_c$ and $\zeta^2 = m_a/m_b$ in which $m_{ab} = \sqrt{m_a m_b}$. In addition, we define $\lambda_{ab} = \sqrt{\lambda_a \lambda_b}$ and $\xi_{ab} = \sqrt{\xi_a \xi_b}$. The magnetic field $\mathbf{H}$ encloses an angle $\theta$ with the $z$ axis, and its projection in the $xy$ plane encloses an angle $\varphi$ with the $x$ axis; see Fig. 1(a).

The anisotropy enters in the GL free energy (1) only through the gauge-invariant gradient term, which becomes isotropic if we choose the following scales of the coordinate axes and vector potential,

$$x = \zeta^{-1/2}\tilde{x}, \quad y = \zeta^{1/2}\tilde{y}, \quad z = \varepsilon\tilde{z}, \quad A_x = \zeta^{1/2}\widetilde{A}_x, \quad A_y = \zeta^{-1/2}\widetilde{A}_y, \quad A_z = \varepsilon^{-1}\widetilde{A}_z, \quad (2)$$

where we denote a quantity $q$ in the scaled isotropic system by $\tilde{q}$. The magnetic flux density $\mathbf{B} = \nabla \times \mathbf{A}$ is then scaled as:

$$B_x = \zeta^{-1/2}\varepsilon^{-1}\widetilde{B}_x, \quad B_y = \zeta^{1/2}\varepsilon^{-1}\widetilde{B}_y, \quad B_z = \widetilde{B}_z. \quad (3)$$

Applying (3) in the free energy expression (1), one finds that the last two terms $f_m = (2\mu_0)^{-1}B^2 - \mathbf{B}\cdot\mathbf{H}$ representing the magnetic energy are transformed into

$$f_m = (2\mu_0)^{-1}(\zeta^{-1}\varepsilon^{-2}\widetilde{B}_x^2 + \zeta\varepsilon^{-2}\widetilde{B}_y^2 + \widetilde{B}_z^2) - (\zeta^{-1/2}\varepsilon^{-1}\widetilde{B}_x H_x + \zeta^{1/2}\varepsilon^{-1}\widetilde{B}_y H_y + \widetilde{B}_z H_z). \quad (4)$$

Note that, the anisotropy is reintroduced in $f_m$, although it vanishes in the gradient term. In general, it is not possible to render both terms in the Gibbs energy isotropic simultaneously. If the superconductor is strongly type II (GL parameter $\kappa \gg 1$) or if the magnetic field are large enough, the magnetic field is nearly uniform on the elementary length scales, and we can adopt a mean-field decoupling scheme, in which we first minimize the magnetic-field energy $f_m$ with respect to $\widetilde{\mathbf{B}}$ and then insert the resulting uniform field back into the free energy [14]. Minimizing the magnetic-field energy $f_m$ (3) with respect to $\widetilde{B}_x$, $\widetilde{B}_y$ and $\widetilde{B}_z$, the applied external magnetic field is scaled in the isotropic system as,

$$H_x = \mu_0^{-1}\zeta^{-1/2}\varepsilon^{-1}\widetilde{B}_x, \quad H_y = \mu_0^{-1}\zeta^{1/2}\varepsilon^{-1}\widetilde{B}_y, \quad H_z = \mu_0^{-1}\widetilde{B}_z. \quad (5)$$

Combing this result with Eq. (3), we find the constitutive relation $\mathbf{B} = \mu_0\mathbf{H}$ in the anisotropic system. Thus, with the aid of $B_x = B\sin\theta\cos\varphi$, $B_y = B\sin\theta\sin\varphi$ and $B_z = B\cos\theta$, in the rescaled isotropic system the magnetic field is related to the original magnetic field as

$$\widetilde{B} = \varepsilon_{\theta\varphi} B, \qquad (6)$$

where $\varepsilon_{\theta\varphi}^{2} = \varepsilon^{2}(\theta,\varphi) = \varepsilon^{2}\sin^{2}\theta(\zeta\cos^{2}\varphi + \zeta^{-1}\sin^{2}\varphi) + \cos^{2}\theta$. In uniaxial superconductors with $\zeta = 1$, $\varepsilon_{\theta\varphi}^{2}$ is reduced to $\varepsilon_{\theta}^{2} = \varepsilon^{2}\sin^{2}\theta + \cos^{2}\theta$ and thus $\widetilde{B} = \varepsilon_{\theta} B$, which coincides with the appropriate result in [14]. In figure 2 we plot the angular dependence of $B$, indicating that the degree of freedom increase to two in biaxial superconductors, and varying the two anisotropy parameters render nonlinear and gradual changes in the profiles of $B(\theta,\varphi)$.

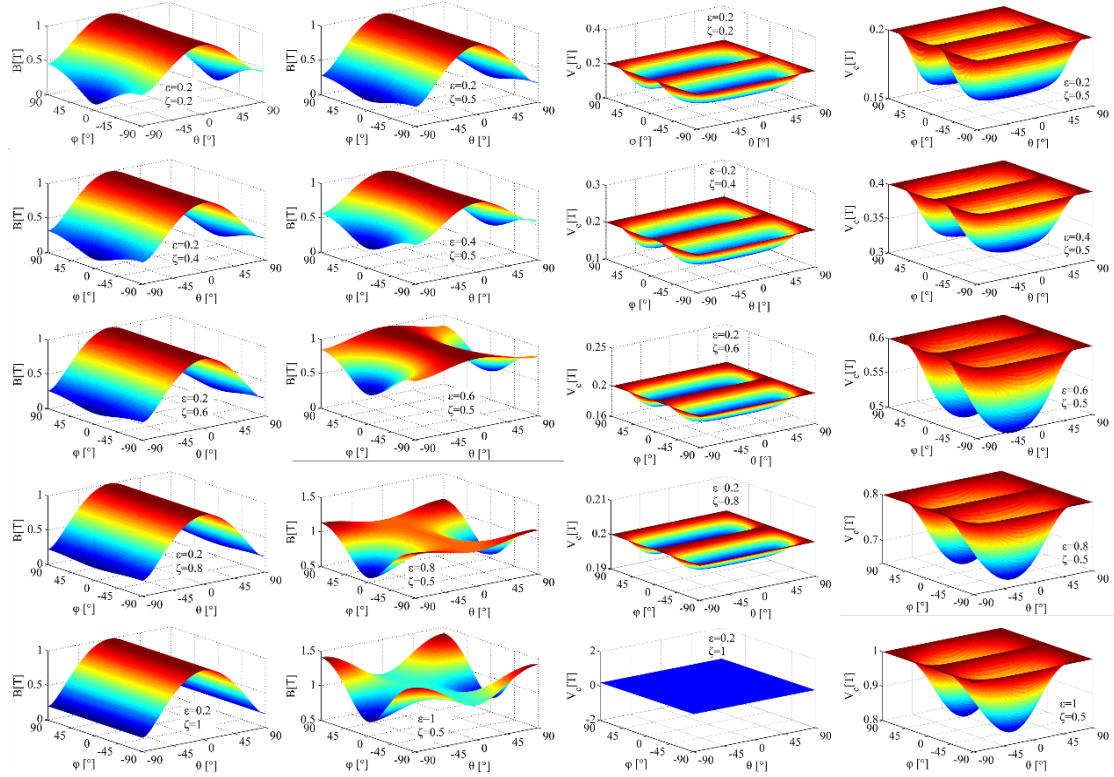

Figure 2. Magnetic field and collective-pinning volume variations with the field direction in an anisotropic biaxial superconductor. The plots are made at different out-of-plane anisotropy $\varepsilon$ and in-plane anisotropy $\zeta$. The scaled isotropic field and volume $\widetilde{B} = 1\mathrm{T}$ and $\tilde{V}_{c} = 1\mathrm{m}^{3}$.

Consider a longitudinal vector $\mathbf{l}_{l} = l_{l}(\sin\theta\cos\varphi, \sin\theta\sin\varphi, \cos\theta)$ directed along the magnetic field $\mathbf{H}$ (along the vortex line). Using Eq. (2) we obtain

$$\tilde{\mathbf{l}}_{l} = (\tilde{l}_{x}, \tilde{l}_{y}, \tilde{l}_{z}) = l_{l}(\zeta^{1/2}\sin\theta\cos\varphi, \zeta^{-1/2}\sin\theta\sin\varphi, \varepsilon^{-1}\cos\theta), \quad l_{l} = \varepsilon\varepsilon_{\theta\varphi}^{-1}\tilde{l}_{l}, \quad \mathbf{l}_{l} \parallel \mathbf{H}. \qquad (7)$$

Consider a "transverse vector" $\mathbf{l}_t = (l_{tx}, l_{ty}, l_{tz})$ with $\hat{l}_{tx} = \cos\theta\cos\varphi\cos\psi - \sin\varphi\sin\psi$, $\hat{l}_{ty} = \cos\theta\sin\varphi\cos\psi + \cos\varphi\sin\psi$ and $\hat{l}_{tz} = -\sin\theta\cos\psi$, which lies in the plane perpendicular to the vortex line. The direction of $\mathbf{l}_t$ is defined by an angle $\psi$. We have the following scaling rules for $\mathbf{l}_t$,

$$\tilde{\mathbf{l}}_t = l_t(\zeta^{1/2}\hat{l}_{tx}, \zeta^{-1/2}\hat{l}_{ty}, \varepsilon^{-1}\hat{l}_{tz}),$$

$$\tilde{l}_t^2 = l_t^2[\zeta(\cos\theta\cos\varphi\cos\psi - \sin\varphi\sin\psi)^2 + \zeta^{-1}(\cos\theta\sin\varphi\cos\psi + \cos\varphi\sin\psi)^2 + \varepsilon^{-2}\sin^2\theta\cos^2\psi],$$

$$\mathbf{l}_t \perp \mathbf{B}. \tag{8}$$

If we denote the radius of the vortex core by $r_c$, the scaled $\tilde{r}_c$ in the isotropic system is then $\tilde{\mathbf{r}}_c = r_c(\zeta^{1/2}\hat{l}_{tx}, \zeta^{-1/2}\hat{l}_{ty}, \varepsilon^{-1}\hat{l}_{tz})$. Since $\tilde{\mathbf{r}}_c$ is not perpendicular to the vortex line direction $\tilde{\mathbf{l}}_l$, $\xi_{ab}$ is given by the projection $r_{c\perp}$ of $\tilde{\mathbf{r}}_c$ on the plane perpendicular to $\tilde{\mathbf{l}}_l$,

$$\xi_{ab} = r_{c\perp} = |\tilde{\mathbf{r}}_c \times \tilde{\mathbf{l}}_l|/|\tilde{\mathbf{l}}_l|. \tag{9}$$

Inserting Eqs. (7) and (8) in Eq. (9), one finds the expression of the radius of the vortex core,

$$r_c(\theta, \varphi, \psi) = \varepsilon_{\theta\varphi}\varepsilon_{\theta\varphi\psi}^{-1}\xi_{ab}, \tag{10}$$

where

$$\varepsilon_{\theta\varphi\psi}^2 = \zeta^{-1}(\sin\varphi\cos\psi + \cos\varphi\sin\psi\cos\theta)^2 + \zeta(\cos\varphi\cos\psi - \sin\varphi\sin\psi\cos\theta)^2 + \varepsilon^2\sin^2\psi\sin^2\theta. \tag{11}$$

Equation (10) can be transformed into $r_c^2(\psi; \theta, \varphi)(\bar{\eta}\cos^2\psi + \bar{\beta}\sin^2\psi) = r_{c0}^2(\theta, \varphi)$, where $r_{c0}^2 = (\varepsilon_{\theta\varphi}\xi_{ab})^2$. The parameters $\bar{\eta}$ and $\bar{\beta}$ are defined as $\bar{\eta} = 0.5[\eta + \beta + (\eta - \beta)\cos 2\psi_1] + \alpha\sin 2\psi_1$ and $\bar{\beta} = 0.5[\eta + \beta - (\eta - \beta)\cos 2\psi_1] - \alpha\sin 2\psi_1$, where $\psi_1 = 0.5\arctan[2\alpha/(\eta - \beta)]$, $\eta(\varphi) = \zeta\cos^2\varphi + \zeta^{-1}\sin^2\varphi$ and $\beta(\varphi, \theta) = \cos^2\theta(\zeta\sin^2\varphi + \zeta^{-1}\cos^2\varphi) + \varepsilon^2\sin^2\theta$ [24]. It is implied that, the shape of the vortex core in an anisotropic biaxial superconductor is an ellipse with the lengths of the axes $r_{c0}\bar{\eta}^{-1/2}$ and $r_{c0}\bar{\beta}^{-1/2}$. The area of the vortex core is then

$$S_{r_c} = \tilde{S}_{r_c}\zeta_{\theta\varphi}, \tag{12}$$

where $\tilde{S}_{r_c} = \pi\xi_{ab}^2$ and $\zeta_{\theta\varphi} = \varepsilon_{\theta\varphi}^2(\bar{\eta}\bar{\beta})^{-1/2}$. In uniaxial superconductors, we have $\zeta_{\theta\varphi} = \varepsilon_\theta$ and $S_{r_c} = \tilde{S}_{r_c}\varepsilon_\theta$, which coincides with the appropriate result in [17, 25].

We now consider the single-vortex collective pinning in anisotropic superconductors. The weak collective pinning for a single vortex interacting with a quenched random potential holds true in superconductors if the elastic energy accumulated along an individual vortex line is significantly larger than the energy of interaction with the other vortices. In isotropic system, comparing the two energies renders $\tilde{L}_c < \tilde{a}_0$, where $\tilde{L}_c$ is the collective pinning length above which the deformation of the vortex exceeds the characteristic length of the collective pinning energy, and $\tilde{a}_0$ is the intervortex space.

Using the scaling law (7), $L_c = \varepsilon \varepsilon_{\theta\varphi}^{-1} \tilde{L}_c$. $\tilde{L}_c = (\tilde{\varepsilon}_0^2 \tilde{\xi}^2 / \pi \tilde{n} \tilde{U}_p^2)^{1/3}$ is the collective pinning length in the isotropic case, with the energy scale $\tilde{\varepsilon}_0 = (\Phi_0/\lambda_{ab})^2 \ln(\lambda_{ab}/\xi_{ab})/(4\pi\mu_0)$ [17]. $\tilde{U}_p$ is the characteristic pinning energy produced by one of the point defects. It is nature to find the volume $V_c$ subjected to the collective pinning,

$$V_c = S_{r_c} L_c = \varepsilon \varepsilon_{\theta\varphi}^{-1} \zeta_{\theta\varphi} \tilde{V}_c, \tag{13}$$

where $\tilde{V}_c = \pi \xi_{ab}^2 \tilde{L}_c$. If the superconductor is uniaxial, then $V_c$ is reduced to $V_c = \varepsilon \tilde{V}_c$ which is consistent with the appropriate one in [14]. In figure 2, we show the profiles of $V_c(\theta,\varphi)$ at different anisotropy parameters. The scaling law for $\tilde{U}_p$ is then $U_p = \varepsilon \varepsilon_{\theta\varphi}^{-1} \zeta_{\theta\varphi} \tilde{U}_p$. If we use $\gamma = \pi n U_p^2$ to represent the degree of disorder, we obtain $\gamma = \varepsilon \varepsilon_{\theta\varphi}^{-1} \zeta_{\theta\varphi} \tilde{\gamma}$. Applying this result in $\tilde{L}_c$, we have $\tilde{L}_c = (\varepsilon \varepsilon_0^2 \xi_{ab}^2 \varepsilon_{\theta\varphi}^{-1} \zeta_{\theta\varphi} \gamma^{-1})^{1/3}$ and

$$L_c(\theta,\varphi) = \varepsilon_{\theta\varphi}^{-1} L_c^c, \tag{14}$$

where $L_c^c = \varepsilon \tilde{L}_c = \varepsilon^{4/3} (\varepsilon_0^2 \xi_{ab}^2 \varepsilon_{\theta\varphi}^{-1} \zeta_{\theta\varphi} \gamma^{-1})^{1/3}$ is the collective pinning energy in the uniaxial superconductor exposed to an applied magnetic field that is directed in the $c$ axis.

The collective pinning energy $U_c = U_p (n S_{r_c} L_c)^{1/2}$ is thus scaled as

$$U_c = \varepsilon_{\theta\varphi}^{-1} \zeta_{\theta\varphi} U_c^c, \tag{15}$$

where $U_c^c = \varepsilon \tilde{U}_c = \varepsilon^{2/3} \varepsilon_0^{1/3} \xi_{ab}^{4/3} \varepsilon_{\theta\varphi}^{1/3} \zeta_{\theta\varphi}^{-1/3} \gamma^{1/3}$. If we denote the direction angle of the pinning force by $\psi_p$, we then obtain the pinning force acting on the vortex line per unit length in this direction,

$$f_p = U_c / L_c r_c = \varepsilon_{\theta\varphi\psi} \bar{f}_p^c, \tag{16}$$

where $\bar{f}_p^c = \varepsilon_{\theta\varphi}^{-1}\zeta_{\theta\varphi}f_p^c$ and $f_p^c = U_c^c(L_c^c\xi_{ab})^{-1} = \varepsilon^{-2/3}\varepsilon_0^{-1/3}\xi_{ab}^{-1/3}\varepsilon_{\theta\varphi}^{2/3}\zeta_{\theta\varphi}^{-2/3}\gamma^{2/3}$ is the pinning force in the uniaxial superconductor with the vortex line directed along the $c$ axis. It is then natural to use the approach in [17, 24] to calculate the critical current density $J_{c\perp}$ in the plane perpendicular to the vortex line, accounting for the misalignment of the directions of the driving force $\mathbf{J}_{c\perp} \times \mathbf{B}$ and the pinning force. We thus obtain $J_{c\perp}$ as

$$J_{c\perp}(\psi_\perp;\theta,\varphi) = \begin{cases} J_{c0}(\sin^2\psi_\perp + \delta^{-1}\cos^2\psi_\perp)^{-1/2}, & \delta \neq 0, \\ J_{c0}, & \delta = 0. \end{cases} \quad (17)$$

Here, $J_{c0} = f_{p0}\Phi_0^{-1}$, $f_{p0}^2 = (\bar{f}_p^c)^2\bar{\eta}$ and $\delta = \bar{\beta}/\bar{\eta}$.

In infinite superconducting slabs or films, the current is constrained in the superconducting $ab$ plane, and the relationship between the critical current density $\mathbf{J}_{c\perp}(\theta,\varphi,\psi_\perp)$ and the in-plane current density $\mathbf{J}(\theta,\varphi,\psi)$ is given by

$$\tan\psi_\perp = \tan(\psi - \varphi)/\cos\theta, \quad J = \Omega J_{c\perp}, \quad (18)$$

where $\Omega = [1 - \cos^2(\psi - \varphi)\sin^2\theta]^{-1/2}$.

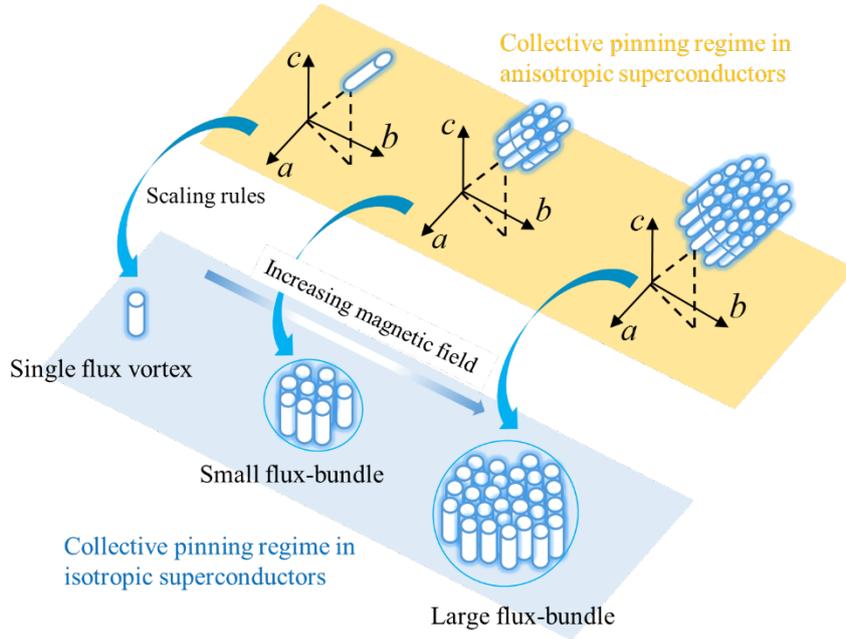

Figure 3. Schematic of the scaling law from anisotropic to isotropic system. Also shown are different collective pinning regimes with increasing magnetic field.

We are now ready to set up the general scaling law for obtaining the desired results of anisotropic superconductors from isotropic systems, Fig. 3. Let us consider an anisotropic biaxial superconductor, characterized by the pinning disorder $\gamma$, London penetration depths ($\lambda_a$, $\lambda_b$ and $\lambda_c$), correlation lengths ($\xi_a$, $\xi_b$ and $\xi_c$) and anisotropy parameters ($\varepsilon$ and $\zeta$), and subjected to an applied external magnetic field **B** enclosing angles $\theta$ and $\varphi$ with respect to the $c$ axis and $a$ axis in the $ab$ plane. The desired quantity $Q$ in the anisotropic superconductor can be obtained from the isotropic result $\tilde{Q}$ through the scaling law:

$$Q\left(\varepsilon,\zeta,\lambda_{a,b,c},\xi_{a,b,c},\gamma,B,\theta,\varphi\right) = s_Q\left(\varepsilon,\zeta,\theta,\varphi\right)\tilde{Q}\left(\tilde{\lambda},\tilde{\xi},\tilde{\gamma},\tilde{B}\right). \tag{19}$$

For the radius of the vortex core, $s_{r_c} = \varepsilon_{\theta\varphi}\varepsilon_{\theta\varphi\psi}^{-1}$; for collective pinning energy, $s_{U_c} = \varepsilon\varepsilon_{\theta\varphi}^{-1}\zeta_{\theta\varphi}$. From the mathematical viewpoint, the scaling process means that a quantity $Q$ in an anisotropic system is decomposed into a prefactor $s_Q$ containing anisotropy parameters and a quantity $\tilde{Q}$ without anisotropy.

## 2.2 Flux-lattice collective pinning in anisotropic superconductors

Different from the single-vortex pinning regime, the elementary scales of lengths and of energy strongly depend on the strength of the magnetic field in flux-lattice collective pinning regime. For vortex line length $L < a_0$, one can neglect the interactions with other vortices, and the displacement field is determined by the competition between the pinning potential and the elasticity of the individual vortex lines. On the other hand, if $L > a_0$, the interactions between vortices become relevant, and one has to construct the three-dimensional physical picture for the vortices and pinning potential; see Fig. 1(b) and Appendix. Applying the scaling law (6) in the intervortex spacing $\tilde{a}_0 = \Phi_0^{1/2}\tilde{B}^{-1/2}$, we find

$$a_0 = \varepsilon_{\theta\varphi}^{1/2}\tilde{a}_0 = \varepsilon_{\theta\varphi}^{1/2}a_0^c, \tag{20}$$

where $a_0^c = \tilde{a}_0 = \Phi_0^{1/2}(\varepsilon_{\theta\varphi}B)^{-1/2}$. Combining Eq. (20) with $\tilde{L}_c = L_c^c\varepsilon^{-1}$, we arrive at

$$\frac{\tilde{L}_c}{\tilde{a}_0} = (\varepsilon_{\theta\varphi}B)^{1/2}\frac{\tilde{L}_c}{\Phi_0^{1/2}} = \frac{\varepsilon_{\theta\varphi}^{1/2}}{\varepsilon}\frac{L_c^c}{a_0} = (\varepsilon_{\theta\varphi}B)^{1/2}\frac{L_c^c}{\varepsilon\Phi_0^{1/2}}. \tag{21}$$

The length scales in the problem of vortex-bundle pinning can grow beyond the London penetration depth $\lambda$, thus the scaling law is thus invalid in certain limited regions of magnetic-field strength. We now consider the small-bundle pinning regime where the magnetic field is moderately large, $\tilde{L}_c[\ln(\lambda/\tilde{L}_c)/\tilde{c}]^{-1/3} < \tilde{a}_0 < \tilde{L}_c$. Making the substitutions $R_c \to R_c \varepsilon_{\theta\varphi}^{-1}\varepsilon_{\theta\varphi\psi}$, $a_0 \to \varepsilon_{\theta\varphi}^{-1/2} a_0$ and $L_c/a_0 \to (\varepsilon_{\theta\varphi}^{1/2}\varepsilon^{-1})(L_c^c/a_0)$ in Eq. (A3), we obtain the characteristic lengths $R_c$ and $L_c^b$ within small-bundle regime,

$$R_c = \varepsilon_{\theta\varphi}\varepsilon_{\theta\varphi\psi}^{-1}R_c^c, \qquad a_0^c < R_c^c < \lambda_{ab}, \tag{22}$$

where $R_c^c = \tilde{R}_c = a_0^c \exp[\tilde{c}(\tilde{L}_c/\tilde{a}_0)^3]$. Equation (22) can be transformed into $R_c^2(\psi;\theta,\varphi)(\bar{\eta}\cos^2\psi + \bar{\beta}\sin^2\psi) = R_{c0}^2(\theta,\varphi)$, where $R_{c0}^2 = (\varepsilon_{\theta\varphi}R_c^c)^2$. This means that, the shape of the flux bundle subjected to the collective pinning is an ellipse with the lengths of the axes $R_{c0}\bar{\eta}^{-1/2}$ and $R_{c0}\bar{\beta}^{-1/2}$. The area of the flux bundle is then $S_{R_c} = S_{R_c}^c \zeta_{\theta\varphi}$, where $S_{R_c}^c = \tilde{S}_{R_c} = \pi(R_c^c)^2$. In uniaxial superconductors, we have $\zeta_{\theta\varphi} = \varepsilon_\theta$ and $S_{R_c} = \varepsilon_\theta S_{R_c}^c$.

Using the scaling law $r_c(\theta,\varphi,\psi) = \xi_{ab}\varepsilon_{\theta\varphi}\varepsilon_{\theta\varphi\psi}^{-1}$, $U_c = \varepsilon\varepsilon_{\theta\varphi}^{-1}\zeta_{\theta\varphi}\tilde{U}_c$ and $L_c^b = \varepsilon\varepsilon_{\theta\varphi}^{-1}\tilde{L}_c^b$, the pinning force $f_p = U_c(r_c L_c^b)^{-1} = \varepsilon_{\theta\varphi\psi}\varepsilon_{\theta\varphi}^{-1}\zeta_{\theta\varphi}\tilde{U}_c\xi_{ab}^{-1}(\tilde{L}_c^b)^{-1} = \varepsilon_{\theta\varphi\psi}\bar{f}_p^c$, where $\bar{f}_p^c = \varepsilon_{\theta\varphi}^{-1}\zeta_{\theta\varphi}f_p^c$ and $f_p^c = (\frac{\tilde{L}_c}{\tilde{a}_0})^2 U_{sv}^c \xi_{ab}^{-1}(L_c^c)^{-1} = (\frac{\tilde{L}_c}{\tilde{a}_0})^2 f_{p,sv}^c$. The expressions $U_{sv}^c = \varepsilon\tilde{U}_{sv}$ and $L_c^c = \varepsilon\tilde{L}_c$ have been used in deriving $f_p^c$. It is implied that the critical current density in the flux lattice is determined as in the single-vortex regime. Equation (17) holds true in this situation, whereas $J_{c0} = f_{p0}B^{-1}S_{R_c}^{-1}$.

Assuming that the magnetic field lies within the $yz$ plane in a uniaxial superconductor (i.e. $\varphi = \pi/2$ and $\zeta = 1$), we find

$$J_{c\perp}(\psi_\perp,\theta,\varepsilon_\theta B) = J_{c0}(\sin^2\psi_\perp + \varepsilon_\theta^{-2}\cos^2\psi_\perp)^{-1/2}, \tag{23}$$

where

$$J_{c0}(\varepsilon_\theta B) = \pi^{-1}f_{p0}(R_c^c)^{-2}(\varepsilon_\theta B)^{-1} = J_{sv}^c(\frac{\tilde{L}_c}{\tilde{a}_0})^2 \exp[-2\tilde{c}(\frac{\tilde{L}_c}{\tilde{a}_0})^3]. \tag{24}$$

If the current $J$ flows in the $xy$ plane at an angle $\psi$, we can calculate $J$ from $J_{c\perp}$ (23) with the help of Eq. (18). In some experiments [5] the samples are rotated maintaining $\mathbf{J} \perp \mathbf{H}$ to obtain a maximum Lorentz force. This corresponding to $\psi = 0$, such that $\psi_\perp = \pi/2$ and $J_{c\perp}(\varepsilon_\theta B) = J_{c0}$. On the other hand, if $\psi = \pi/2$ then $\psi_\perp = 0$ and $J_{c\perp}(\theta, \varepsilon_\theta B) = J_{c0}\varepsilon_\theta$. It should be noted that in general cases the mass anisotropy ratio $\varepsilon^2 = m_{ab}/m_c \ll 1$, and the estimation $\varepsilon_\theta \approx \cos\theta$ renders a unified formula for the two cases, $J_{c\perp} = J_{sv}^c (\frac{\tilde{L}_c}{\tilde{a}_0})^2 \exp[-2\tilde{c}(\frac{\tilde{L}_c}{\tilde{a}_0})^3]$ where $J_{sv}^c = f_{p,sv}^c \Phi_0^{-1}$.

If we keep the magnetic field being increased, the length scales of the collective-pinning flux bundle may exceed the London penetration depth. It may occur in the large-bundle regime where the magnetic field is considerable large, $\tilde{a}_0 < \tilde{L}_c [\ln(\lambda_{ab}/\tilde{L}_c)/\tilde{c}]^{-1/3}$. If we assume the scaling law being also valid in large-bundle pinning regime (although this assumption may cause discrepancy from the real situation), we obtain a unified formula of the critical current density within a wide range of the magnetic field,

$$j_c(\tilde{B}_\sigma) = \begin{cases} 1, & \tilde{B}_\sigma < 1 \\ \tilde{B}_\sigma^2 \exp(-2\tilde{c}\tilde{B}_\sigma^3), & 1 < \tilde{B}_\sigma < [\ln(\lambda_{ab}/\tilde{L}_c)/\tilde{c}]^{-1/3} \\ \left(\varepsilon^{-1} L_c^c \lambda_{ab}^{-1}\right)^2 \tilde{B}_\sigma^{-6}, & \tilde{B}_\sigma > [\ln(\lambda_{ab}/\tilde{L}_c)/\tilde{c}]^{-1/3} \end{cases} \quad (25)$$

where $\tilde{B}_\sigma = \tilde{L}_c/\tilde{a}_0 = \sigma \tilde{B}^{1/2}$, the pinning-strength parameter $\sigma = \tilde{L}_c \Phi_0^{-1/2} \propto \gamma^{-1/3}$, the scaled magnetic field $\tilde{B} = \varepsilon_\theta B$ and the normalized critical current density $j_c = J_{c\perp}/J_{sv}^c$.

Within the small-bundle pinning region, the decrease in the critical current density $j_c$ with the increasing scaled magnetic induction $\tilde{B}$ shows an exponential dependence, $j_c \propto \exp(-\tilde{B}^{3/2})$. Whereas in the large-bundle pinning region, this dependence changes into a monomial $j_c \propto \tilde{B}^{-3}$. In fact, the single-vortex pinning holds true more probably at $\tilde{B}_\sigma \ll 1$, and we thus introduce a crossover $\tilde{B}_{\sigma 1}$ $(0 < \tilde{B}_{\sigma 1} \ll 1)$ from the single-vortex pinning regime to the small-bundle pinning regime. The crossover to the large-bundle region is $\tilde{B}_{\sigma 2} \sim [\ln(\lambda_{ab}/\varepsilon^{-1} L_c^c)/\tilde{c}]^{-1/3}$. Considering the continuity at the crossover of the adjacent regions, one finds

$$j_c(\tilde{B}_\sigma) = \begin{cases} 1, & \tilde{B}_\sigma < \tilde{B}_{\sigma 1} \\ \exp(2\tilde{c}\tilde{B}_{\sigma 1}^3 - 2\tilde{c}\tilde{B}_\sigma^3), & \tilde{B}_{\sigma 1} < \tilde{B}_\sigma < \tilde{B}_{\sigma 2} \\ \tilde{B}_{\sigma 2}^6 [\exp(2\tilde{c}\tilde{B}_{\sigma 1}^3 - 2\tilde{c}\tilde{B}_{\sigma 2}^3)]\tilde{B}_\sigma^{-6}, & \tilde{B}_\sigma > \tilde{B}_{\sigma 2} \end{cases} \quad (26)$$

The scaling laws for the flux-lattice pinning are now established. In Fig. 3 we schematically present the general idea of the scaling approach within different collective pinning regions. Through Eq. (26), one can determine $j_c$ in an anisotropic uniaxial superconductor exposed to magnetic fields with any magnitude and direction. Since $j_c$ depends on $\tilde{B}_\sigma \propto (\varepsilon_\theta B)^{1/2}$, the plots of $j_c(B)$ at various $\theta$ are essentially collapsed onto a single curve $j_c(\varepsilon_\theta B)$. This scaling rule has been highlighted in recent experiments [20]. We give the detailed formulas of the scaling law, which are of importance in understanding the underlying physics of the anisotropic superconductor and explaining the experimental phenomenon from the essential mechanism. In the next section we will apply the scaling rules to explore the fundamentals of the field and angular dependence of $j_c$ in anisotropic superconductors, where possible we discuss the relevance of the theory to the experimental results.

**3. Field and angle dependence of critical current density in anisotropic superconductors**

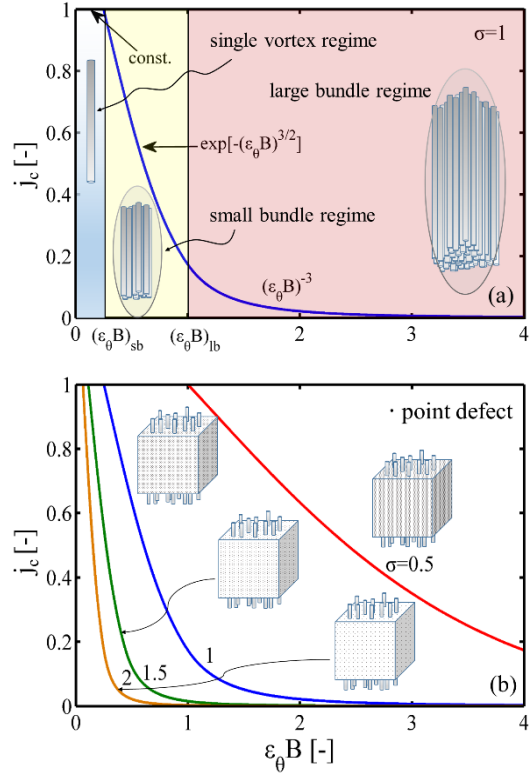

Figure 4. Normalized critical current density variation with scaled magnetic induction $\tilde{B} = \varepsilon_\theta B$. (a) Collective pinning regime. Within $\tilde{B} < \tilde{B}_{sb}$ a superconductor enters into single-vortex pinning regime. $\tilde{B}_{sb} < \tilde{B} < \tilde{B}_{lb}$ and $\tilde{B} > \tilde{B}_{lb}$ correspond to small-bundle and large-bundle collective pinning region, with $j_{c\perp} \propto \exp(-\tilde{B}^{3/2})$ and $j_{c\perp} \propto \tilde{B}^{-3}$. (b) A set of samples with different pinning-strength parameter. The smallest pinning-strength parameter represents the largest defect density or unit pinning energy.

In Fig. 4(a) we plot $j_c(\tilde{B})$ in different collective pinning. $j_c$ is constant from 0 to $\tilde{B}_{sb}$, exponentially dropping from $\tilde{B}_{sb}$ to $\tilde{B}_{lb}$ and algebraically dropping when crossing over $\tilde{B}_{lb}$. The key point is that according to the present scaling rules, within the weak collective pinning regime $j_c$ depends on $B$ and $\theta$ only through $\tilde{B}$. The definition of the variable $\tilde{B}$ is correct even in some cases of strong pinning [5]. It has been recently realized in experiment [20], .. curves at different $\theta$ collapsing onto one curve $J_c(\varepsilon_\theta B)$. Figure 4(b) presents $j_{c\perp}(\tilde{B})$ for different pinning. Note that the theory is limited to treat weak collective pinning, so $j_{c\perp}(\tilde{B})$ is a representative result for vortices in weak pinning potential.

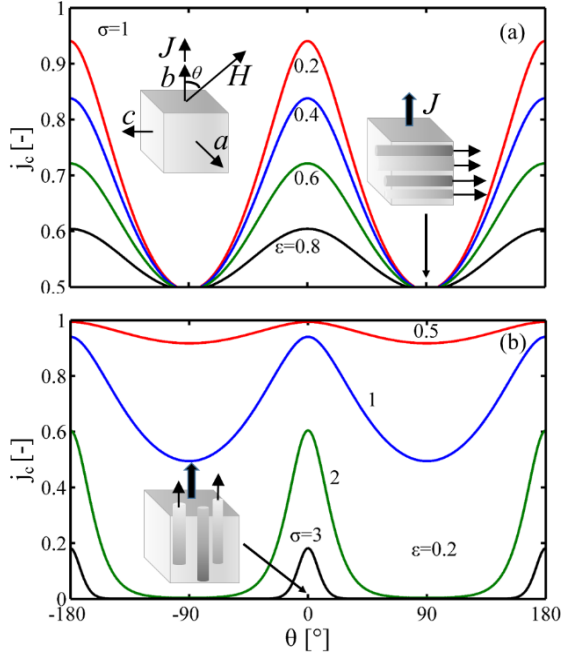

Figure 5. Normalized critical current density $j_c$ versus field angle $\theta$ at $B = 0.5\text{T}$. (a) Variation of $j_c(\theta)$ with mass anisotropy ratio $\varepsilon$ for a superconductor with pinning-strength parameter $\sigma = 1[\text{T}^{-1/2}]$. (b) A set of superconductors with different $\sigma$ at $\varepsilon = 0.2$. The inserts indicates the orientation of vortexes and current in the frame of the crystal principal axes.

Based upon the scaling rules, we can consider the general concept of anisotropic weak-collective-pinning in the isotropic system. In the weak random pinning potential, the vortex lattice cannot be held by individual pins, since the averaging pinning force over individual pins is null. Thus, fluctuations in the pinning potential over a certain volume deform the elastic vortex lines by a small distortion. The elasticity of the vortex is largely described by the line tension for single vortex, whereas by lattice elastic constant for flux lattice. On the other hand, the collective pinning potential per unit volume is not related to the pinning energy of one pin, but to the fluctuations in the pinning energy over unit volume. Taken together, these are quantitatively described by $j_{c\perp}(\tilde{B})$ at different pinning-strength parameters $\sigma$, and $j_{c\perp}(\tilde{B})$ depends on a combination of the defect density and unit pinning energy since $\sigma \propto \gamma^{-1/3} \propto (nU_p^2)^{-1/3}$. The weak collective pinning renders a decrease in the critical current density with increasing field.

The strong pinning interaction shows a softening in the elastic moduli of the flux lattice with the increasing magnetic field [26]. The strong pinning defects pin the lattice individually, thus the pinning energy is linear with the defect density $n$. Within the strong pinning regime, a second maximum occurs in $j_{c\perp}(\tilde{B})$ curve at high fields [20], named as the "fishtail effect".

The angles $\theta$, $\varphi$ and $\psi$ determining the directions of the magnetic field and current are the additional degrees of freedom in anisotropic superconductors. As for the uniaxial superconductors with the field fixed in the $ab$ plane, the remaining is angle $\theta$ of the field direction. Figure 5(a) presents a set of superconductors with different mass anisotropy ratios $\varepsilon$ at $B = 0.5$T and $\sigma = 1$T$^{-1/2}$. The vortices aligned with the $c$ axis in the $\varepsilon$ =0.2, 0.4, 0.6 and 0.8 samples result in minima of $J_c$, whereas a maximum is found wherever the vortices are perpendicular with the $c$ axis. The main features of $j_c(\theta)$ are also found in the experiment [6], the uniaxial superconductor mimicking the experimental superconducting trilayer device where a weak-pinning layer is sandwiched by two strong-pinning layers. The scaling formulas suggest that varying $\theta$ cause a change in the scaled field $\tilde{B} = \varepsilon_\theta B$ at constant $B$. Therefore at $\theta = \pi/2$ (corresponding to $\theta = 0$ in Fig. 4), under the simplification $\varepsilon^2 \ll 1$ one simply finds a roughly maximum $\varepsilon_\theta$. Recalling that $j_{c\perp}(\tilde{B})$ decrease with $\tilde{B}$, this scenario then predicts a maximum in $j_c$. Furthermore, the superconductors with smaller $\varepsilon$ show a noticeably raising $j_c(\theta)$, which means the anisotropy enhances the critical current density irrespective of the alignment degree of the vortices with respect to the crystal principle axes. This is a favorable effect in view of technological applications of the new HTS materials. Figure 5(b) presents $j_c(\theta)$ for a set of superconductors with different pinning-strength parameter $\sigma$ at the fixed $\varepsilon$. The main point is that at any direction of the magnetic field, a smaller $\sigma$ corresponds to a higher critical current density. So we conclude that the more defective nature (simultaneously increasing the defect density and unit pinning strength) of the uniaxial superconductor turns into an advantage as it results in higher $J_c$.

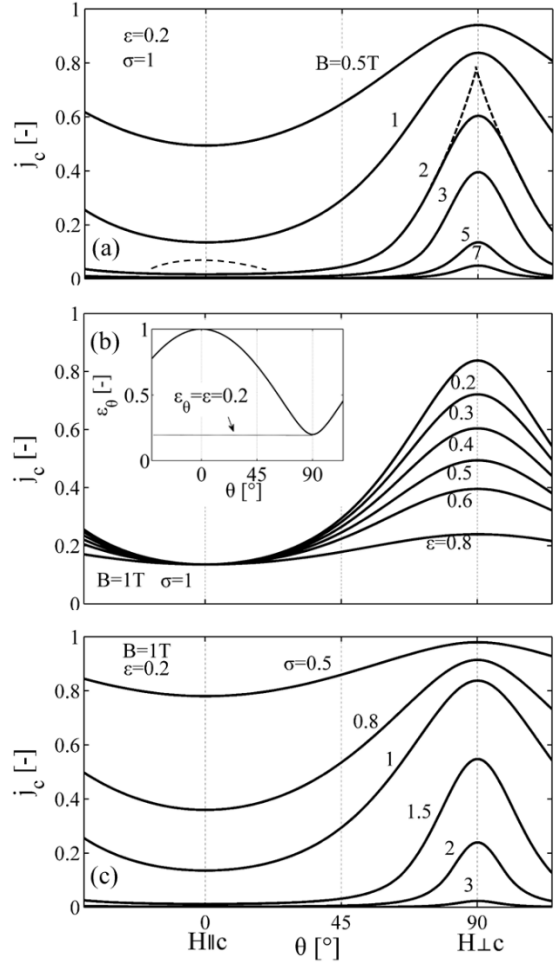

Figure 6. Angular dependence of $j_c$ at several (a) magnetic fields, (b) mass anisotropy parameters and (c) pinning strength parameters. The dashed line in (a) is an estimation of $j_c(\theta)$ at 2T in strong pinning regime, compared to the results obtained from the weak collective pinning theory.

In Figs. 6(a), (b) and (c) we plot $j_c(\theta)$ for different magnetic fields, mass anisotropy parameters and pinning strength parameters, respectively. In the weak collective pinning regime, $j_c$ is a decreasing function of the scaled field $\varepsilon_\theta B$. $\varepsilon_\theta(\theta)$ shows a broad maximum centered at $\theta = 0°$ and a relatively sharp minimum at $\theta = 90^{\circ}$ [see the insert in Fig. 6(b)]. So, at constant fields $j_c(\theta)$ exhibits a broad minimum centered at $\mathbf{H} \parallel ab$, and a sharp maximum centered at $\mathbf{H} \parallel c$. When crossover from weak pinning to strong pinning, a sharp $ab$-plane peak and broad $c$-axis peak appear in $j_c(\theta)$, as shown in the dashed curve in Fig. 6(a), according to the fishtail effect. This behavior is reminiscent of the experimental results [5] on YBa$_2$Cu$_3$O$_7$ films. In Fig. 6(b) a knot can be found at $\mathbf{H} \parallel c$ for curves at different $\varepsilon$. This is simply explained by the vanish of $\varepsilon$ in $\varepsilon_\theta$ for $\theta = 0^{\circ}$; the essence is that when the vortices are aligned with the $c$ axis, the energy of the vortices and the interaction with the defects cannot be influenced by the anisotropy. It is indicated from Fig. 6(c) that the more defective nature of the superconductor renders a higher critical current density. Interestingly, it also reduces the gradient of $J_c(\theta)$, which is another merit in view of applications.

## 4. Conclusions

We develop the scaling laws for anisotropic biaxial superconductors within different collective-pinning regimes. The scaling approach provides a simple way to investigate the complex physics of the anisotropic superconductors, also it serves as a predictive tool for the general behaviors accounting for anisotropy. Using the scaling rules, we obtain a unified formula for the field dependence of the critical current density, coinciding with the recent findings on weak-pinning superconductors. Strong pinning may lead to some corrections to the results of weak-collective pinning. We give a new insight into the pinning mechanism, the defect density and unit pinning energy jointly affecting the collective pinning energy. We obtain the angular dependence of the critical current density in a set of superconductors, and the results are well explained when going some deep into the scaling laws.

## 5. Acknowledgments

This work was performed with supports from the National Natural Science Foundation of China (11372120, 11032006 and 11421062), National Key Project of Magneto-Restriction Fusion Energy Development Program (2013GB110002) and Fundamental Research Funds for the Central Universities (lzujbky-2014-227).

**Appendix: Vortex-line bundle and vortex-lattice pinning in isotropic system**

The following derivations, essentially based upon [14], give the descriptions for the flux bundle within the weak collective pinning regime in isotropic superconductors. In the absence of the external force field, the vortex lattice adjusts itself to the underlying disorder potential via shear and tilt deformations alone. We denote the longitudinal dimension (along the vortex line) by $L_c^b$, and the transverse dimension (perpendicular to the vortex line) by $R$, Fig. 1(b). The competition between the tilt energy $c_{44}[u(L_c^b)^{-1}]^2$ and the shear energy $c_{66}(uR^{-1})^2$ determines the aspect of the vortex bundle,

$$\frac{L_c^b}{R_c} = (\frac{c_{44}}{c_{66}})^{1/2} > 1, \tag{A1}$$

If deformation occurs within a transverse length $a_0 < R_c < \lambda$ (dispersive regime), the tilt modulus $c_{44} \simeq c_{44}^0 (R_c \lambda^{-1})^2$ in which $c_{44}^0$ is for the uniform tilt. Whereas at large distances $R_c > \lambda$ (nondispersive regime), the tilt modulus $c_{44} \simeq c_{44}^0$. Thus, Eq. (A1) is rewritten as

$$L_c^b = \begin{cases} \dfrac{R_c^2}{a_0}, & a_0 < R_c < \lambda, \\ \dfrac{\lambda}{a_0} R_c, & R_c > \lambda, \end{cases} \tag{A2}$$

Now, we consider the pinning energy $E_{pin}$ due to energy fluctuations in the disorder potential; for a deformation with amplitude $u < \xi$, the estimation for $E_{pin}$ is $E_{pin} = \dfrac{(V\Delta)^{1/2}}{\xi} \dfrac{u}{\xi}$. $V = R_c^2 L_c^b$ is the volume, and $\Delta$ is the disorder parameter describing energy fluctuations in the disorder potential. Balancing the elastic shear energy within the volume $V$ against the pinning energy, one finds $u = \dfrac{(L\Delta)^{1/2}}{c_{66}\xi^2} \dfrac{R_c}{L_c^b}$. Applying Eq. (A2) in this result, we arrive at the displacement field $u(R, L_c^b)$. This is valid within a regime bounded by the conditions $u(R_c, L_c^b) = \xi$. We are now ready to set down

$$R_c \simeq \begin{cases} a_0 \exp[\tilde{c}(\dfrac{L_c}{a_0})^3], & a_0 < R_c < \lambda, \\ \lambda(\dfrac{L_c}{a_0})^3, & R_c > \lambda, \end{cases} \tag{A3}$$

Here, $L_c$ is the single-vortex collective pinning length, $a_0 = (\Phi_0/B)^{1/2}$ is the intervortex space. The collective pinning lengths $R_c$ and $L_c^b$ determine the real-space boundaries of the small vortex bundle of small spatial fluctuations $u \leq \xi$. Within the volume characterized by the length scales $R_c$ and $L_c^b$, a maximum displacement $u$; $\xi$ is accumulated due to elastic deformation produced by the pinning potential, and thus the vortex lattice is collectively pinned at a single metastable state.

For a relaxed vortex lattice, the elastic shear energy, tilt energy and the collective pinning energy are equal. Within the collective pinning volume $V_c = R_c^2 L_c^b$, a displacement $u \approx \xi$ on a scale $R_c$ produces the shear energy $c_{66}(\xi/R_c)^2 V_c$, which can express the basic energy scale. Applying Eqs. (A2) and (A3) in this result, we obtain the collective pinning energy

$$U_c \simeq \begin{cases} U_{sv} \simeq \varepsilon_0 \xi^2 L_c^{-1}, & a_0 > L_c \\ U_{sv} \dfrac{L_c}{a_0} e^{2\tilde{c}(L_c/a_0)^3}, & L_c[\ln(\lambda/L_c)/\tilde{c}]^{-1/3} < a_0 < L_c, \\ U_{sv} (\dfrac{\lambda}{a_0})^2 (\dfrac{L_c}{a_0})^4, & a_0 < L_c[\ln(\lambda/L_c)/\tilde{c}]^{-1/3}, \end{cases} \quad (A4)$$

where $U_{sv}$ is the collective pinning energy for single-vortex pinning regime. The critical current density $J_c$ is obtained by balancing the driving Lorentz force $J_c B R_c^2$ against the pinning force $U_c(\xi L_c^b)^{-1}$, i.e. $J_c = U_c(B\xi R_c^2 L_c^b)^{-1}$.

Inserting Eqs. (A2) and (A3) for the small-bundle and large-bundle transverse dimension $R_c$, we obtain the field-dependent critical current density

$$J_c \simeq \begin{cases} J_{sv} = U_{sv}(\Phi_0 \xi L_c)^{-1}, & a_0 > L_c \\ J_{sv} (\dfrac{L_c}{a_0})^2 e^{-2\tilde{c}(L_c/a_0)^3}, & L_c[\ln(\lambda/L_c)/\tilde{c}]^{-1/3} < a_0 < L_c, \\ J_{sv} (\dfrac{a_0}{\lambda})^2 (\dfrac{a_0}{L_c})^4, & a_0 < L_c[\ln(\lambda/L_c)/\tilde{c}]^{-1/3}, \end{cases} \quad (A5)$$